\documentclass[final]{aipproc}

\layoutstyle{6x9}

\begin{document}

\title{ Effective interactions and long distance symmetries in the
  Nucleon-Nucleon system}
  \classification{03.65.Nk,11.10.Gh,13.75.Cs,21.30.Fe,21.45.+v}  \keywords{NN Interaction, Renormalization, Effective Interactions, Symmetries, Large $N_c$} \author{ \underline{E. Ruiz
    Arriola}~\footnote{Speaker at {\em
      Chiral10 WORKSHOP}, 21-24 Jun 2010, Valencia, Spain } \, and A. Calle Cord\'on}{ address={Departamento de
      F\'{\i}sica At\'omica, Molecular y Nuclear \\ Universidad de
      Granada \\ E-18071 Granada, Spain.}  }

\begin{abstract}
Effective interactions, when defined in a coarse grained sense, such
as $V_{\rm low k}$ at the scale $\Lambda = 450 {\rm MeV}$, display a
remarkable symmetry pattern. Serber symmetry works with high accuracy
for spin triplet states.  Wigner SU(4) spin-isospin symmetry with
nucleons in the fundamental representation works only for even partial
waves exactly as predicted by large $N_c$ limit of QCD with accuracy
${\cal O} (1/N_c^2)$. This suggests tailoring the very definition of
effective interactions to provide a best possible fulfillment of long
distance symmetries. With the $V_{\rm low k}$ definition Wigner
symmetry requires that chiral potentials have low cut-offs
$\Lambda_\chi \sim 450 {\rm MeV}$.  Perturbative saturation of
exchanged heavy mesonic resonances does not faithfully display the
Serber symmetry pattern.
\end{abstract}

\maketitle

\section{Introduction} 

The NN interaction is too hard to be treated directly in {\it ab
  initio} nuclear structure calculations for medium and heavy nuclei
mainly because the strong short range repulsion generates short
distance correlations. While much work has been devoted to handle this
problem, one should recognize that calculations become involved
precisely in the region where the interaction is least known since NN
elastic scattering experiments probe distances larger than the minimal
de Broglie wavelength $1 /\sqrt{m_\pi M_N} \sim 0.5 {\rm fm}$
corresponding to pion production threshold. The momentum space $V_{\rm
  low k}$ approach~\cite{Holt:2003rj} (see
e.g. Ref.~\cite{Bogner:2009bt} for a review) takes a Wilsonian point
of view of integrating out high energy components and working within
an effective Hilbert space. The out-coming potentials are smooth and
more amenable to mean field and perturbative treatments at the expense
of introducing scale dependent three- and higher-body forces. For the
two-body problem one replaces~\cite{Holt:2003rj} 
the Lippmann-Schwinger (half-off shell) equation for the {\it bare}
potential
\begin{eqnarray}
T (k',k;k^2) = V (k',k) + \frac{2}{\pi}\int_0^\infty \, 
\frac{dq \, q^2}{k^2-q^2} V ( k' , q) T ( q ,k;k^2) 
\end{eqnarray}
by the equivalent $V_{\rm low k}$ potential defined by the equation restricted to $(k,k') \le \Lambda$, 
\begin{eqnarray}
T (k',k;k^2) = V_{\rm low k} (k',k) + \frac{2}{\pi}\int_0^\Lambda \, 
\frac{dq \, q^2}{k^2-q^2} V_{\rm low k} ( k' , q) T (q,k;k^2) 
\label{eq:vlowk}
\end{eqnarray}
For instance, in perturbation theory the  $V_{\rm low
  k}$ potential becomes
\begin{eqnarray}
V_{\rm low k} ( k' , k) = V ( k' , k)  
+ \frac{2}{\pi} \int_\Lambda^\infty \,  \frac{dq \, q^2}{k^2-q^2}
V ( k', q) V ( q , k)+ \dots 
\label{eq:vlowk-pert}
\end{eqnarray}
The amazing finding~\cite{Holt:2003rj} was that {\it all} high
precision interactions, i.e. fitting the NN elastic scattering data
with $\chi^2 /{\rm DOF} \sim 1$ and including One Pion Exchange (OPE),
have quite different momentum space $V(k,k')$ behaviour, but reduced
to an {\it universal} $ V_{\rm low k}$ potential when $\Lambda \le 2.1
{\rm fm}^{-1}$. The coordinate space equivalent (alias $V_{\rm High R}$)
corresponds to the universality of boundary conditions obtained from
integrating in from large distances taking the experimental phase
shifts {\it and} the OPE potential down to a shortest possible
distance cut-off $r_c$ where {\it all other} components of the
potential are
negligible~\cite{CalleCordon:2008cz,CalleCordon:2009ps,RuizArriola:2009bg}.

\section{Long distance symmetries}

In this contribution we point out that such an interpretation unveils
important symmetries of the effective NN interaction relevant to
Nuclear Structure. Given a symmetry group with a generic element $G$,
a standard symmetry means that $[ V,G] =0 $ implies $[ V_{\rm low
    k},G] =0 $. The reverse, however, is not true.  We define a {\it
  long distance symmetry} as a symmetry of the effective interaction,
i.e. $[ V_{\rm low k},G] =0 $ but $[ V ,G] \neq 0 $. From a
renormalization viewpoint that corresponds to a symmetry of the
potential broken only by counter-terms. In coordinate space the
symmetry is broken by the boundary condition at the cut-off radius,
$r_c$~\cite{CalleCordon:2008cz,CalleCordon:2009ps,RuizArriola:2009bg}.
Important features of the NN interaction are both the spin-orbit
interaction and the tensor force. However, it was recognized in early
partial wave studies (see e.g. \cite{Holinde:1975vg}) that there
appear strong statistical correlations among different channels which could
be better handled by defining suitable linear combinations for phase
shifts at fixed orbital angular momentum.  This may
be understood by separating the NN potential as the sum of central
components and (small) non-central components,
\begin{eqnarray} 
{\cal V}_{NN} = V_0 + V_1 \, , 
\end{eqnarray}
where $[ \vec L, V_0] =0 $ whereas $[ \vec J, V_1] = 0 $ and $[ \vec
  L, V_1] \neq 0 $. 
 The zeroth order potential commutes with $L,S,T$ and so the
corresponding phase shift is denoted as $\delta_L^{ST}$.  The total
potential commutes with the total angular momentum $J = L+S$ and the
phase-shift is $\delta_{LJ}^{ST}$.  Using first order perturbation
theory around the central potential one obtains
\begin{eqnarray}
\delta_{LJ}^{ST} 
 &=& \delta_C^{LST} + \delta_{S,1}\ \delta_T^{LST} (S_{12}^{J})_{LL}  
+ \delta_{LS}^{LST} \frac12 \left[ J(J+1) - L(L+1) - S(S+1) \right]\, ,
\label{eq:delta-STLJ}
\end{eqnarray}  
where $ (S_{12}^{J})_{J-1,J-1} = -2(J-1)/(2J+1)$, $ (S_{12}^{J})_{J,J}
= 2 $, $ (S_{12}^{J})_{J+1,J+1} = -2(J+2)/(2J+1)$.  The phases
$\delta_C^{LST}$, $\delta_T^{LST}$ and $\delta_{LS}^{LST}$ represent
the center of the multiplet, the splitting due to spin-orbit and
the tensor force and can be obtained by inverting
Eq.~(\ref{eq:delta-STLJ}).

\begin{figure}[ttt]
\begin{tabular}{ccc}
\centerline{
\includegraphics[height=4cm,width=4cm,angle=270]{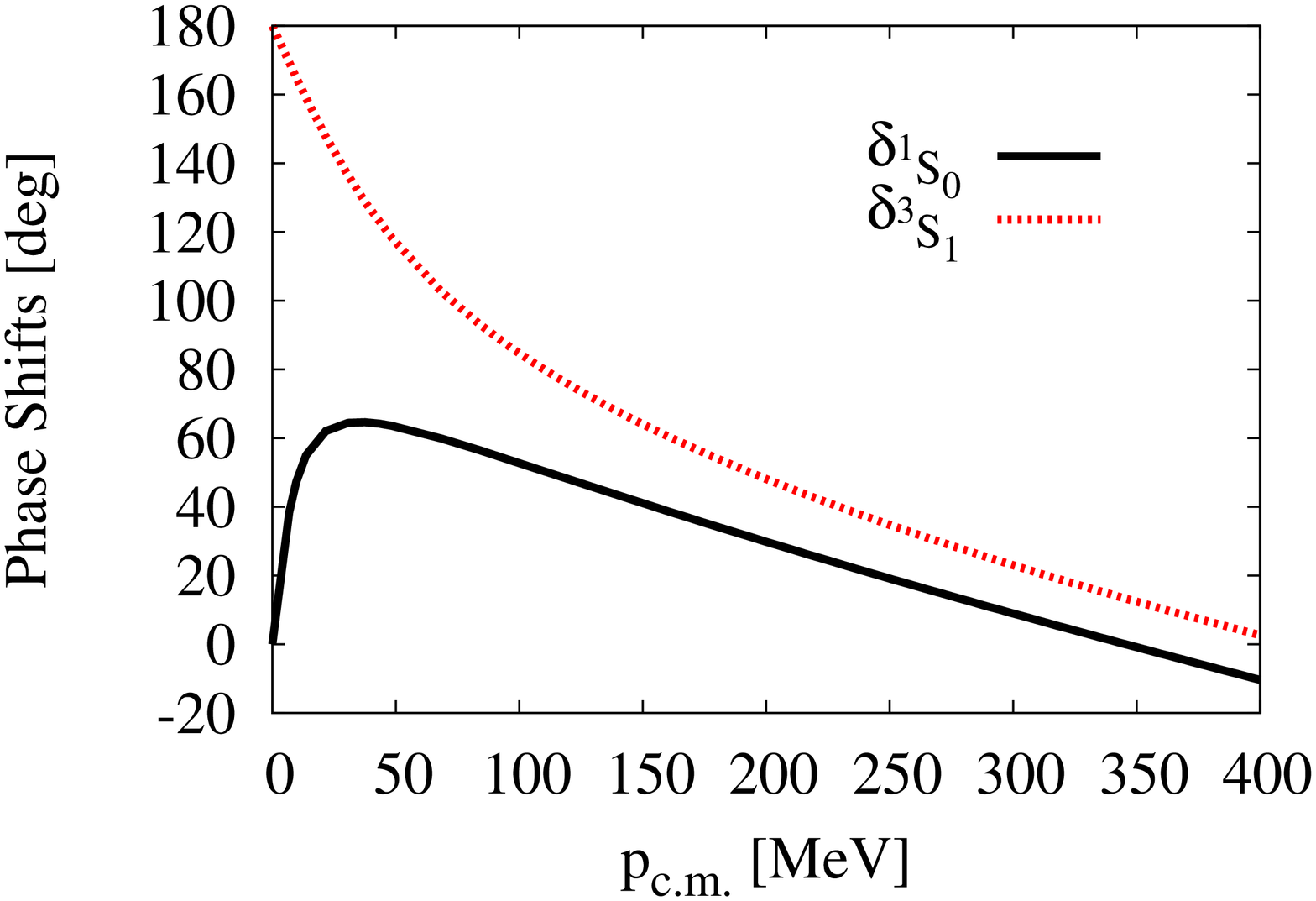} 
\includegraphics[height=4cm,width=4cm,angle=270]{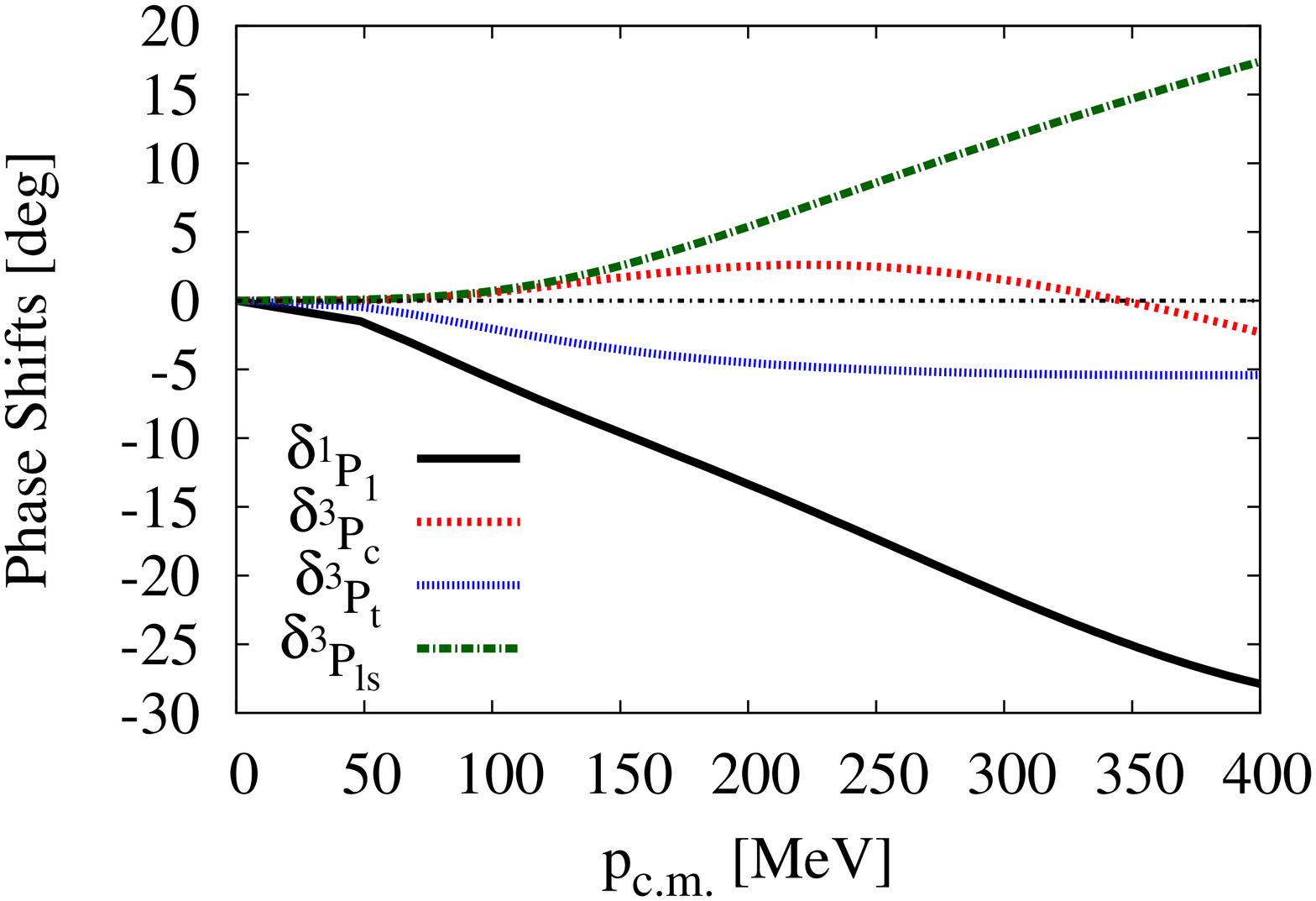} 
\includegraphics[height=4cm,width=4cm,angle=270]{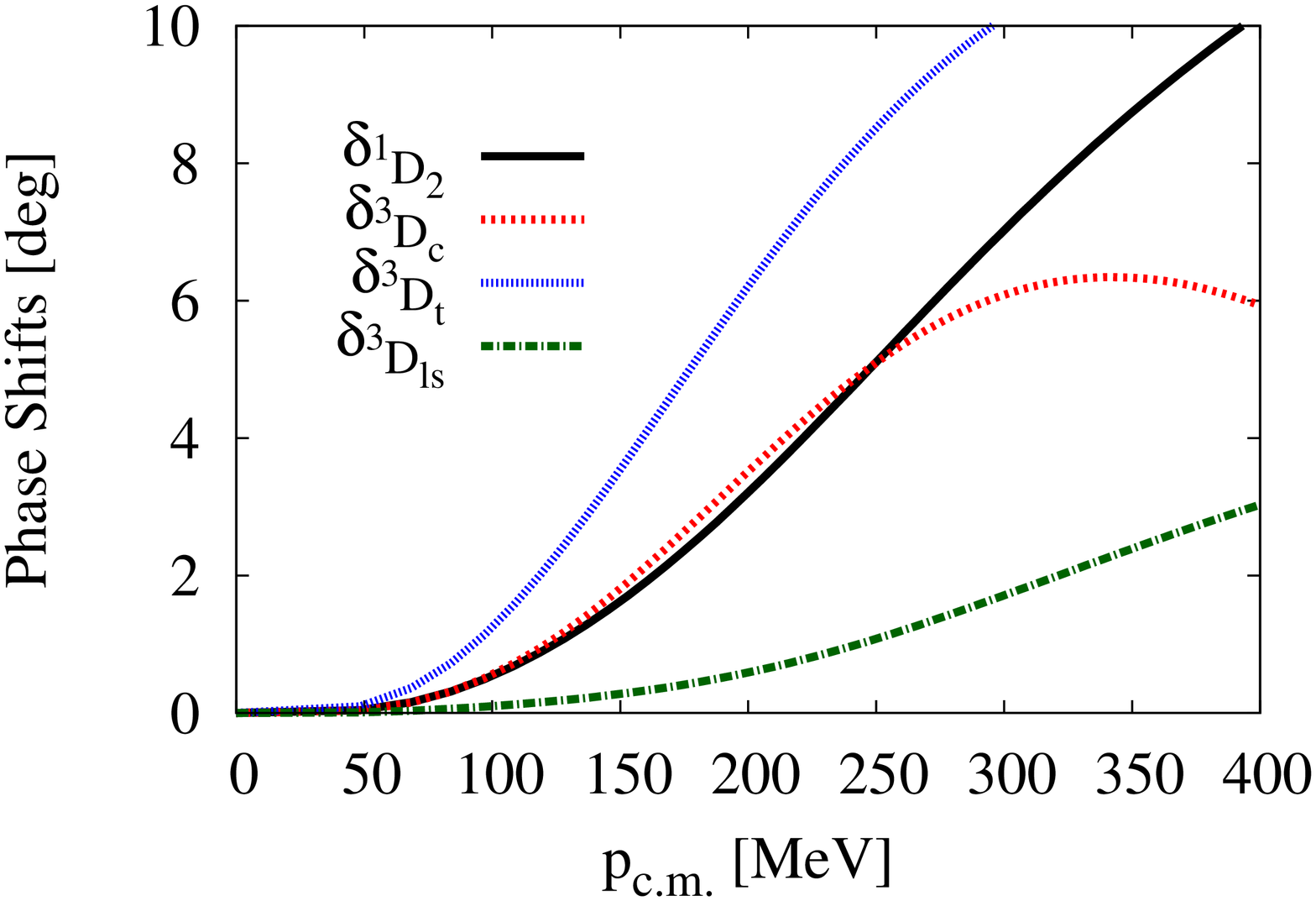}  
\includegraphics[height=4cm,width=4cm,angle=270]{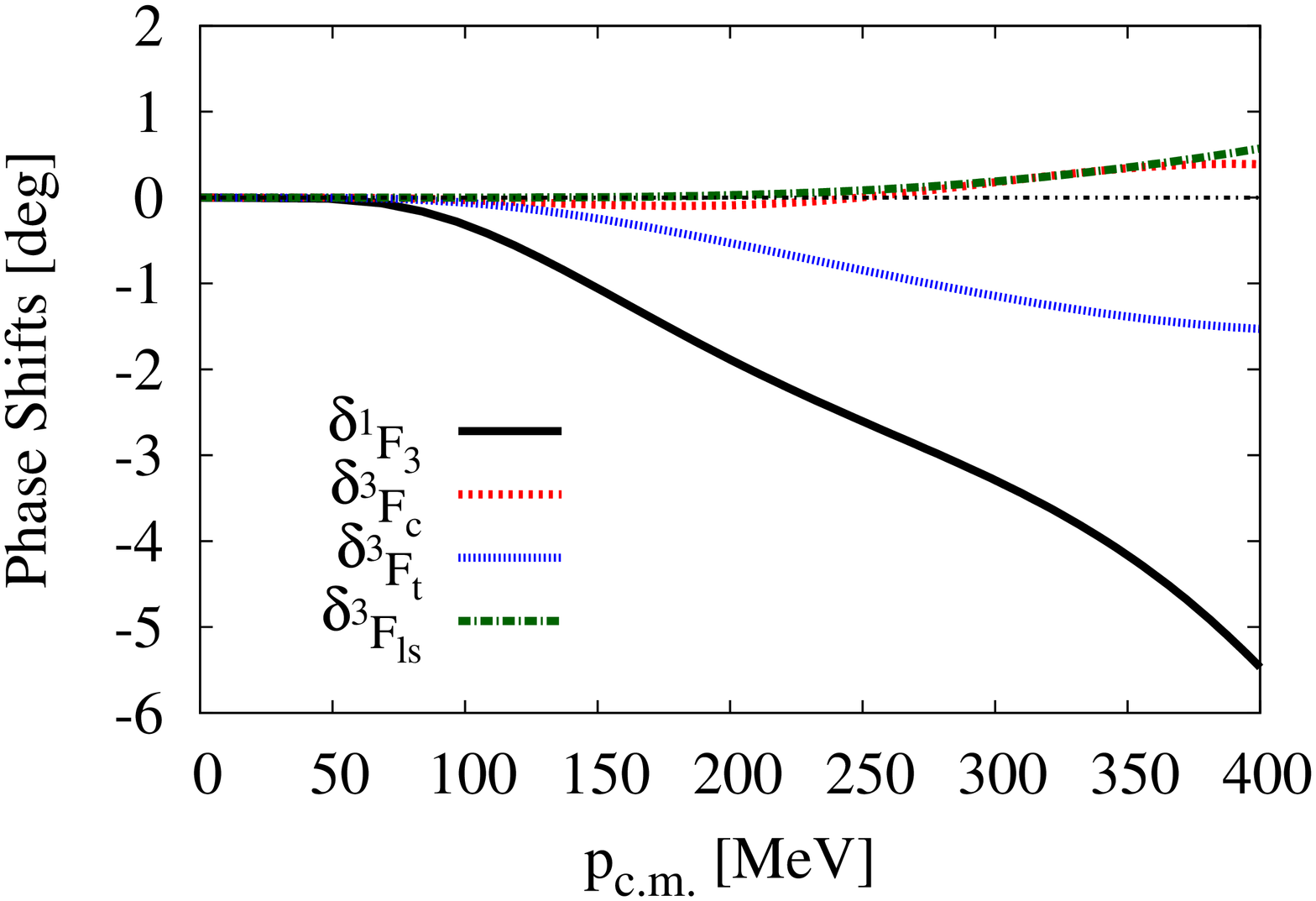}}
\\
\centerline{
\includegraphics[height=4cm,width=4cm,angle=270]{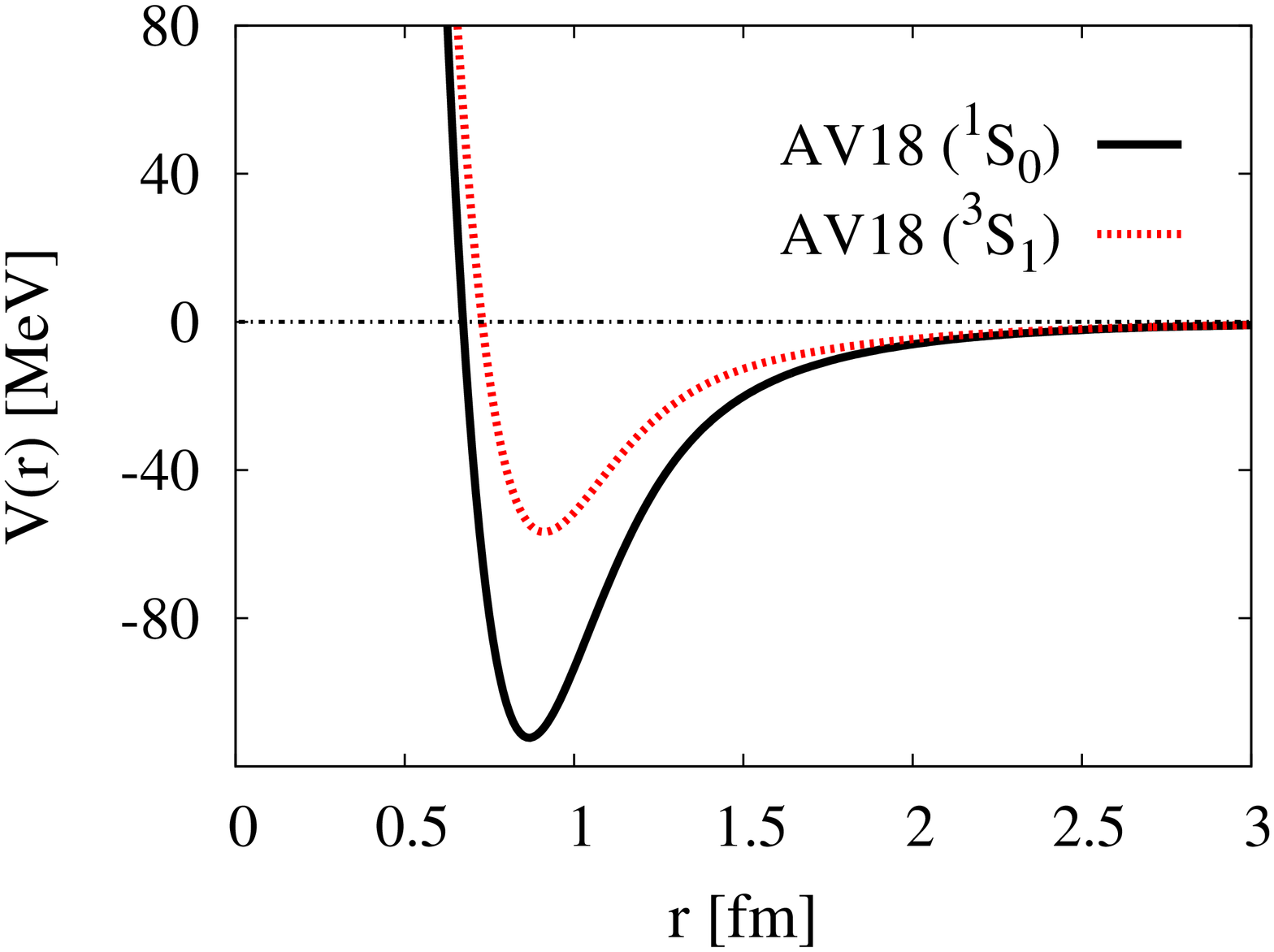}
\includegraphics[height=4cm,width=4cm,angle=270]{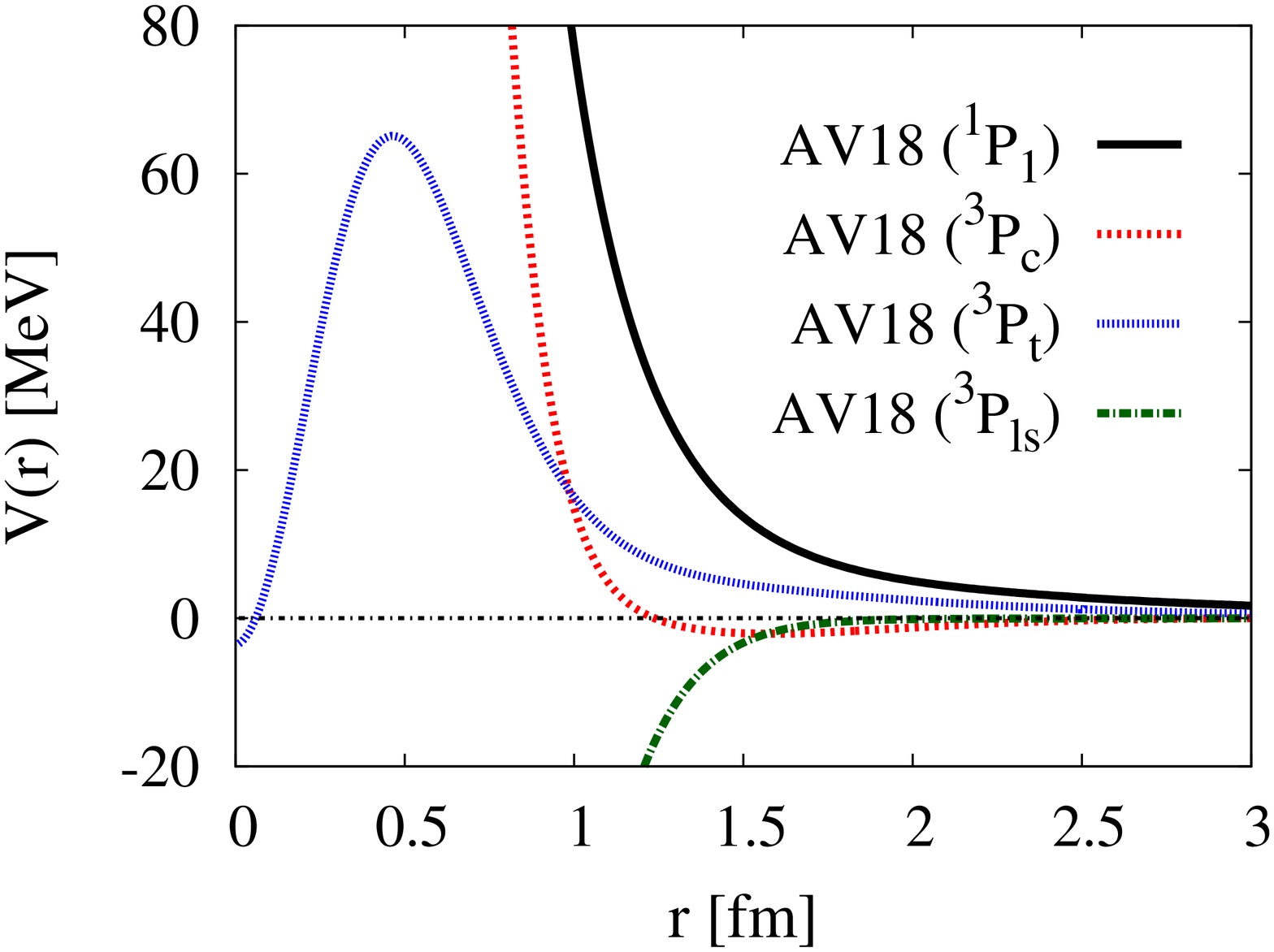}
\includegraphics[height=4cm,width=4cm,angle=270]{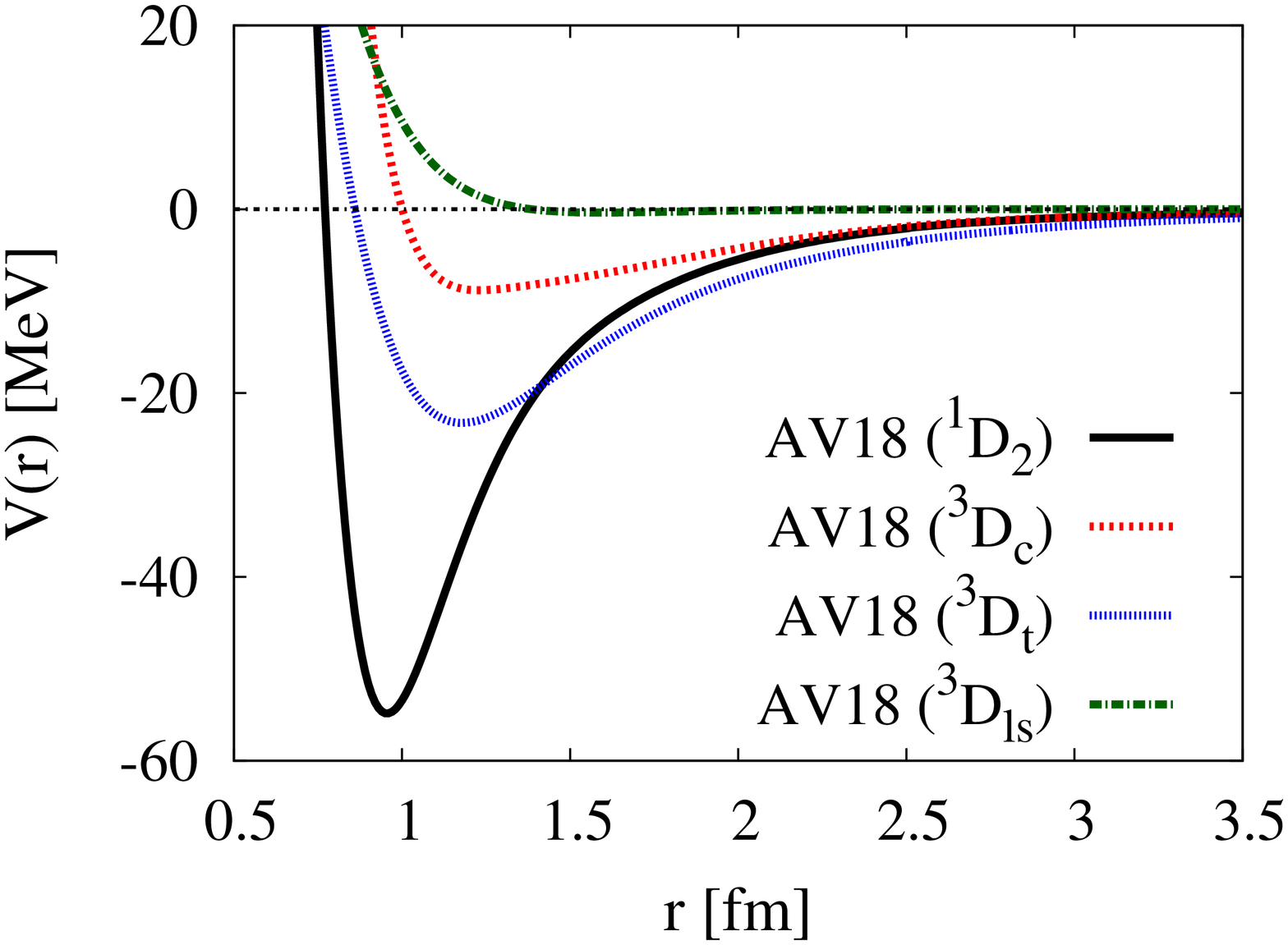}
\includegraphics[height=4cm,width=4cm,angle=270]{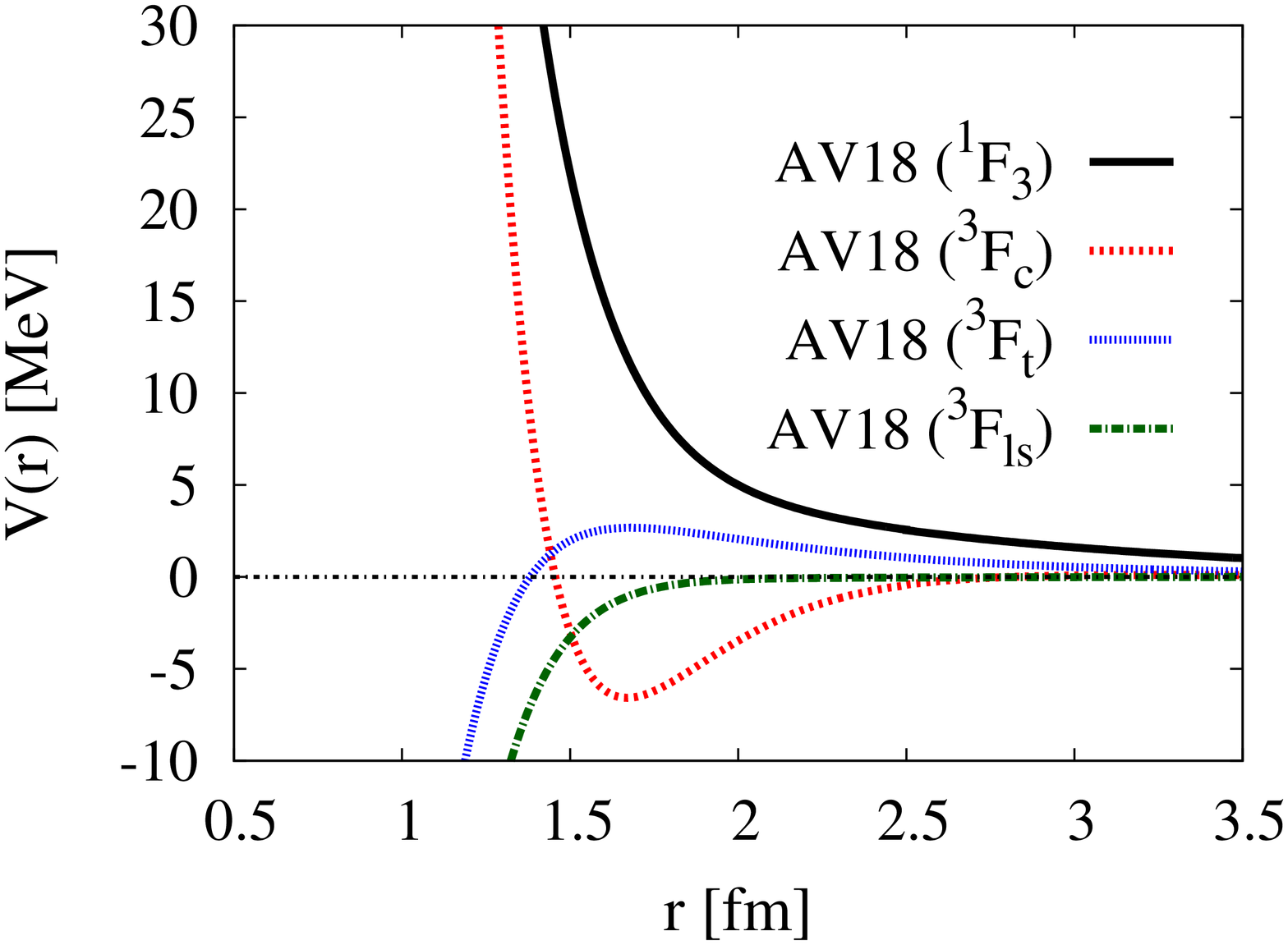}
}
\\
\centerline{
\includegraphics[height=4cm,width=4cm,angle=270]{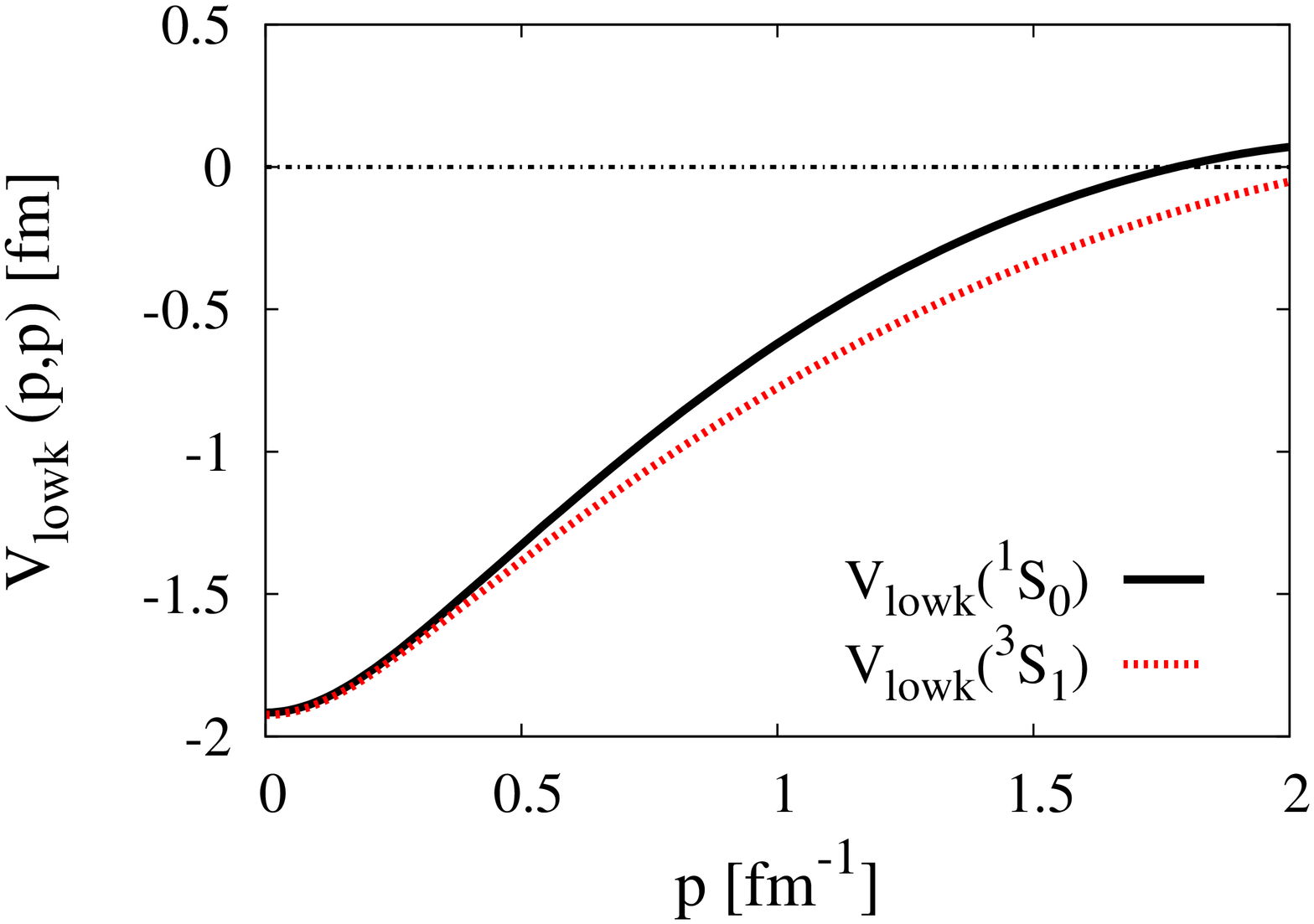}
\includegraphics[height=4cm,width=4cm,angle=270]{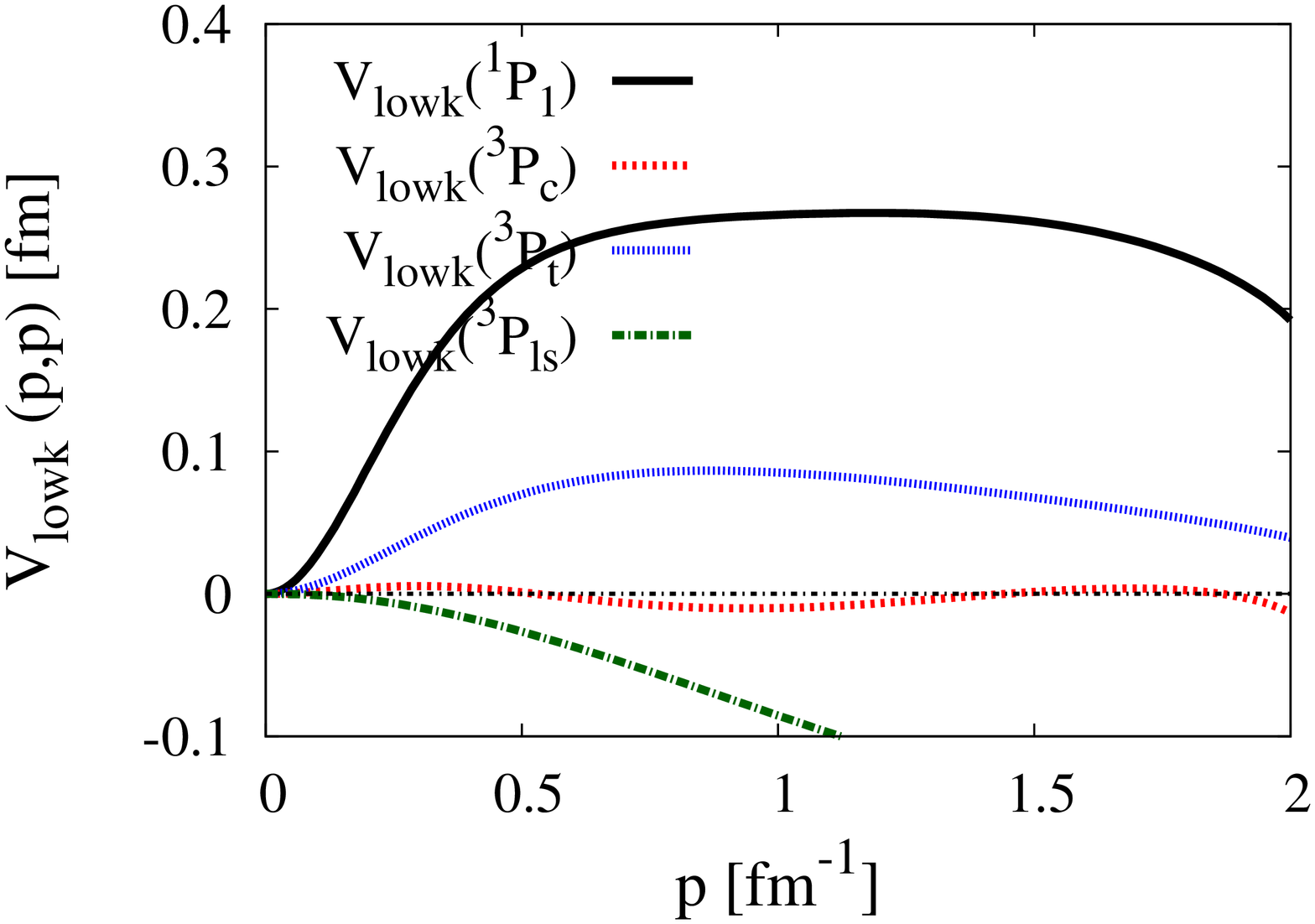}
\includegraphics[height=4cm,width=4cm,angle=270]{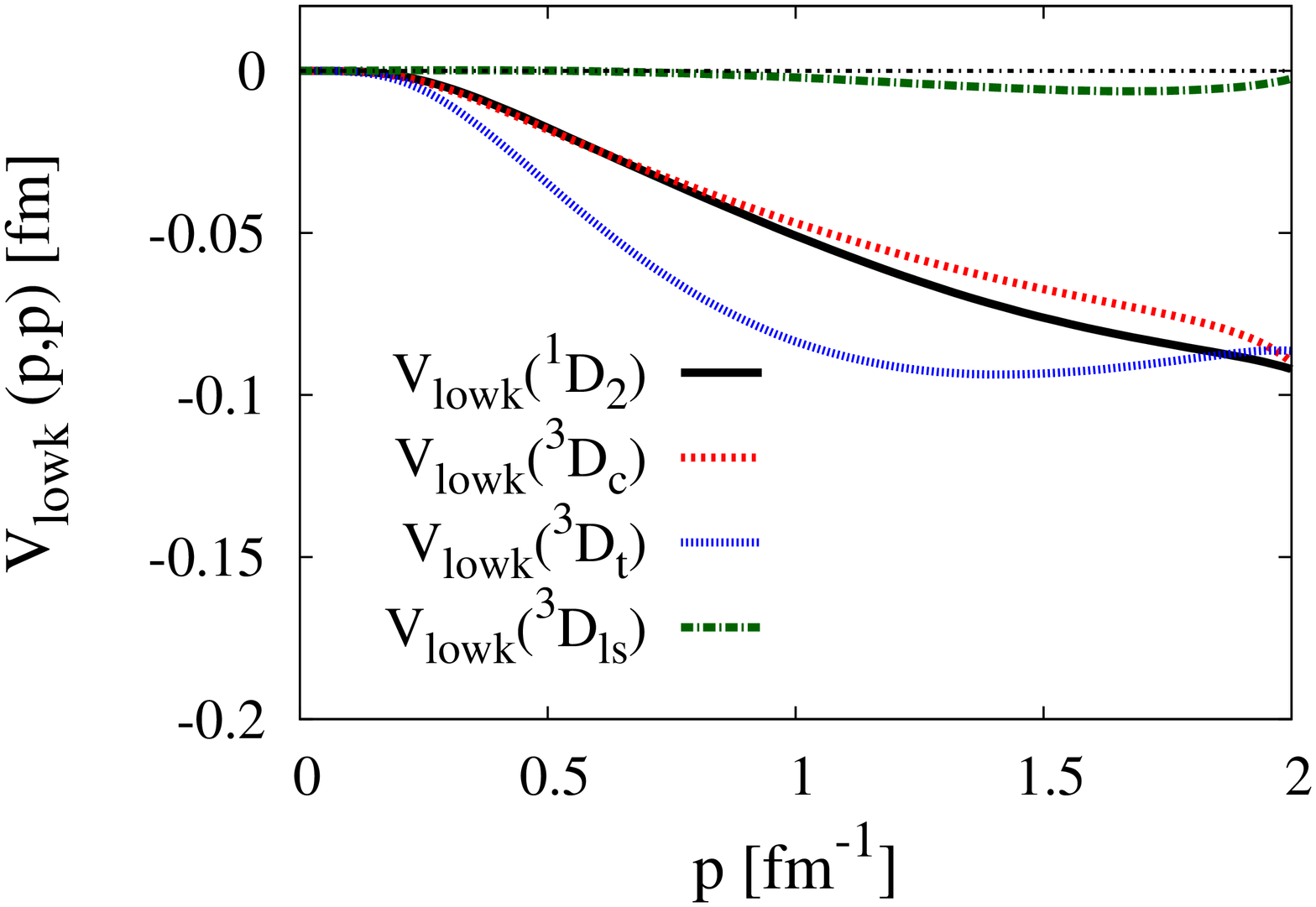}
\includegraphics[height=4cm,width=4cm,angle=270]{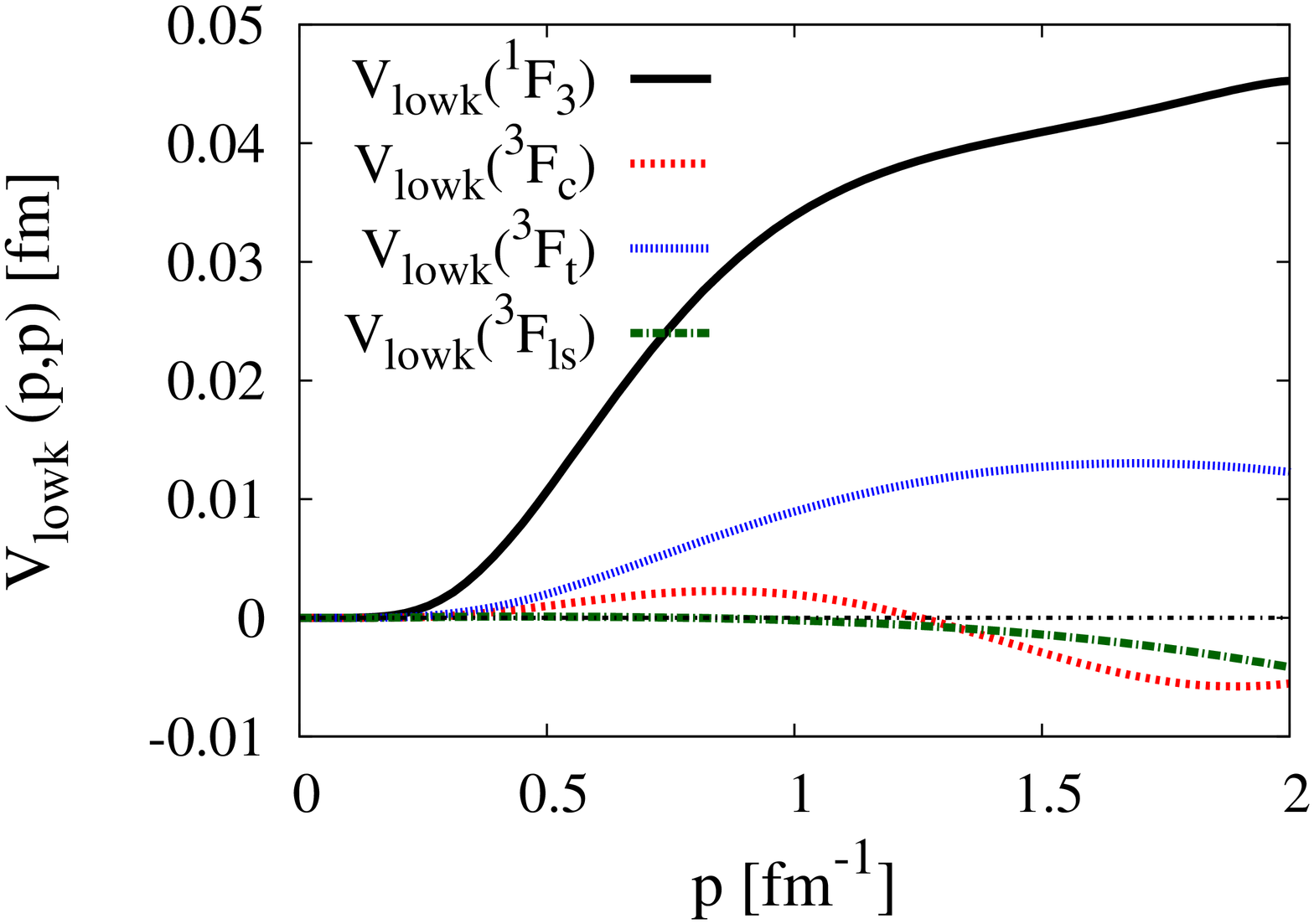}}
\end{tabular}
\caption{Top panel:Average values of the phase
  shifts~\cite{Stoks:1993tb} (in degrees) as a function of the CM
  momentum (in MeV). Middle panel:  Argonne V-18
  potentials~\cite{Wiringa:1994wb} (in MeV) as a function of distance
  (in fm). Bottom panel: 
Diagonal $V_{\rm low k} (p,p)$
  potentials (in fm) as a function of the momentum p (in ${\rm
    fm}^{-1}$)~\cite{Holt:2003rj}.}
\label{fig:wigner-serber}
\end{figure}

In Fig.~\ref{fig:wigner-serber} we show the low partial waves with
orbital angular momentum $L=0,1,2,3$ using the average values of the
phase shifts analyzed by the Nijmegen group~\cite{Stoks:1993tb}.
Likewise we have depicted also similar combinations for the
Argonne-V18 potential~\cite{Wiringa:1994wb} and the corresponding
$V_{\rm low k}$ potential~\cite{Holt:2003rj} obtained from it. Spin
singlet ($^1S_0$) and spin triplet ($^3S_1$) S-waves have very
different phase shifts. However, both the low energy interaction as
well as the potential at long distances are very similar. This
property also holds for D- and G-waves, i.e., $^1S_0=^3S_1$, $^1D_2 =
^3D_c$ and $^1G_3 = ^3G_c$ and corresponds to spin independence of the
forces in even-L channels (Wigner symmetry).  On the contrary, triplet
P-waves and F-waves average to null, i.e.  $^3P_c = ^3F_c= 0$ in
agreement with the Serber symmetry $| f_{pn} (\theta) |^2 = |f_{pn}
(\pi-\theta)|^2$ in the $pn$ differential cross section in the CM
scattering angle, $\theta$. All this could be observed in old
analyses~ \cite{Holinde:1975vg}. The new feature unveiled in
Refs.~\cite{CalleCordon:2008cz,CalleCordon:2009ps,RuizArriola:2009bg}
has been the recognition that this is a property of the effective
interaction. This amazing fact suggests that the $V_{\rm low k}$
approach is a good filter for an otherwise unforeseen symmetry. Note
also that LS couplings in D- and F-waves are rather small within this
framework, i.e. $^3D_{LS}=^3F_{LS}=0$.  In passing we note that Wigner
symmetry requires that $V_{\rm low k}$ Chiral forces to
N$^3$LO~\cite{Entem:2003ft} have their cut-off $\Lambda_\chi \sim
\Lambda_{\rm low k} $~\cite{CalleCordon:2009ps} hindering larger
$\Lambda_\chi \sim 600 {\rm MeV} $ values where increasing violations
are found.

\section{LARGE $ N_c$}

While QCD is entitled to eventually generate all features of Nuclear
Physics, one may profit from a semiquantitative analysis based on
distinct properties of the fundamental theory.  Some time
ago~\cite{Kaplan:1996rk}, it was found that within a large $N_c$
expansion the leading piece of the NN potential is ${\cal O} ( N_c)$
and has the tensorial structure
\begin{eqnarray}
V_{NN} (\vec x)  = V_c (r) + \tau_1 \cdot \tau_2 \left[ \sigma_1 \cdot \sigma_2 W_S (r) 
  + S_{12}(\hat x)  W_T (r) \right] + {\cal O} (1/N_c)
\label{eq:VNN-largeNc}
\end{eqnarray}
with corrections (comprising relativistic corrections, spin orbit,
meson widths, etc.), suppressed by a relative $1/N_c^2$. This
potential {\it only } complies to the Wigner symmetry for {\it even}
partial waves. These large-$N_c$ counting rules are based on quark and
gluon dynamics, but for large distances quark-hadron duality allows to
saturate them by the standard (multi-)meson exchange
picture~\cite{Banerjee:2001js}. The striking
thing~\cite{CalleCordon:2008cz,CalleCordon:2009ps,RuizArriola:2009bg}
is that this symmetry pattern emerges in the effective interaction
!!. This important point prevents us from using large $N_c$ literally,
but rather as a long distance symmetry. This way, the ubiquitous fine
tunings, triggered by unknown short distance physics, are efficiently
disentangled from long distance physics with the help of
renormalization.  We are pursuing this large-$N_c$ framework for
Nuclear Physics with rather encouraging results~\cite{Cordon:2009pj}
for the low partial waves and the deuteron in the case of One Boson
Exchange (OBE) and its Meson Exchange Currents~\cite{Cordon:2010ez} where only
$\pi,\sigma,\rho,\omega,A_1$ contribute to
Eq.~(\ref{eq:VNN-largeNc})~\cite{Banerjee:2001js}.

\section{Resonance saturation vs Scale Saturation}

The momentum space $V_{\rm low k}$ approach~\cite{Holt:2003rj} makes
clear that the long distance behaviour is not determined by the low
momentum components of the original potential {\it only}; one has to
add virtual high energy states which also contribute to the
interaction at low energies. Actually, in the limit $\Lambda \to 0$
one may Taylor expand the effective S-wave interaction
\begin{eqnarray}
V_{\rm low k}(k,k') = (M_N /16 \pi^2)\left[ C_0 + C_2 (k^2 +
  k'^2) + \dots \right]
\end{eqnarray}
Plugging this ansatz into Eq.~(\ref{eq:vlowk}) and fixing the $T$
matrix at low energies to the scattering length $\alpha_0$, effective
range $r_0$, one gets (for $C_2=0$) a result depending on the scale
$\Lambda$,
\begin{eqnarray}
C_{0} (\Lambda) = (16 \pi^2 \alpha_{0}/M_N)/ (1 - 2 \alpha_{0} \Lambda/\pi) 
\label{eq:C0}
\end{eqnarray} 
which for $\Lambda \sim 150 (200) {\rm MeV}$ reproduces the exact
$V_{\rm low k}$ value in the $^1S_0 (^3S_1)$ channel.  The inclusion
of tensor force mixing and range corrections saturates to almost
$100\%$ smoothly the $V_{\rm low k}$ for the same $\Lambda \sim 250
{\rm MeV}$ and complies to the Wigner symmetry pattern, $C_{^1S_0} =
C_{^3S_1}$ as can be seen in Fig.~\ref{fig:wigner-serber} at
$p=0$. Higher partial waves and the relation to Skyrme forces are
analyzed further in Ref.~\cite{Arriola:2010hj}. Note that since $M_N \sim
N_c$ for $C_0 \sim N_c$ one needs $\alpha_0 \sim N_c^2$ and $\Lambda
\sim 1/N_c^2$ in Eq.~(\ref{eq:C0}). 

Inspired by the success of the resonance saturation hypothesis of the
exchange forces in $\pi\pi$ scattering (see e.g. \cite{Ecker:1988te}),
the OBE picture~\cite{Machleidt:1987hj} low momentum
contributions from the exchange of heavier mesons have been identified
as generating the counter-terms for chiral
potentials~\cite{Epelbaum:2001fm}. Taking e.g. a heavy
scalar meson with mass $m \gg \Lambda$ and coupling $g$ one can use
Eq.~(\ref{eq:vlowk-pert}) and the potential reads
\begin{eqnarray}
\frac{g^2}{({\bf p'}-{\bf p}
)^2+m^2} = \frac{g^2}{m^2} - \frac{ g^2 ({\bf p'}-{\bf p}
)^2}{m^4} + \dots  = C_0 + C_2 \left({\bf p}^2+
     {\bf p}'^2 \right) + C_1 {\bf p} \cdot {\bf p}' + \dots \, , 
\label{eq:low-q}
\end{eqnarray} 
Note that here $C_0 = g^2 /m^2 $ does not depend on $\Lambda$, unlike
Eq.~(\ref{eq:C0}). The OBE resonance matching to chiral potentials
besides identifying terms scaling differently in $N_c$, $C_0^{\rm OBE}
\sim N_c$ vs $C_0^{\rm Chiral} \sim g_A^4/f_\pi^2 \sim N_c^2 $ do not
comply to Serber symmetry for P-waves as it happens for the $V_{\rm
  low k}$ determination~\cite{CalleCordon:2009ps}. This does not mean,
however, that the effective interaction cannot be represented in the
polynomial form of Eq.~(\ref{eq:low-q}), but rather that the
coefficients cannot be computed {\it directly} and generically as the
Fourier components of the potential, since the corrections are not
necessarily small (unlike the $\pi\pi$ case).

\section{Conclusions}

The non-trivial fact that the Wigner and Serber symmetries occur
unequivocally for $ V_{\rm low k} $ and not so much for the bare $V$
reinforces the $ V_{\rm low k} $-approach as an efficient and
symmetry-based coarse grained interaction. Once the symmetries emerge,
simplifications are expected, and this affects the very definition of
effective interactions.

\begin{theacknowledgments}
Supported by the Spanish DGI and FEDER funds with grant
FIS2008-01143/FIS, Junta de Andaluc{\'\i}a grant FQM225-05, and EU
Integrated Infrastructure Initiative Hadron Physics Project contract
RII3-CT-2004-506078.
\end{theacknowledgments}

\bibliographystyle{aipproc}   


\end{document}